# Embedded Network Test-Bed for Validating Real-Time Control Algorithms to Ensure Optimal Time Domain Performance

Ayan Mukherjee, Anindya Pakhira, Saptarshi Das, Indranil Pan, Amitava Gupta

*Abstract*—**The paper presents a Stateflow based network test-bed to validate real-time optimal control algorithms. Genetic Algorithm (GA) based time domain performance index minimization is attempted for tuning of PI controller to handle a balanced lag and delay type First Order Plus Time Delay (FOPTD) process over network. The tuning performance is validated on a real-time communication network with artificially simulated stochastic delay, packet loss and out-of-order packets characterizing the network.**

## I. Introduction

NETWORKED control system (NCS) has emerged as a new interdisciplinary research domain with the introduction of real-time network in place of the traditional closed loop control systems. Within a short span of time NCS has caught the attention of wide range of communities across the engineering research spectrum. This popularity can be attributed to inherent advantages that NCS enjoys over the traditional control system as well as its multi-disciplinary origin (communication, control, instrumentation, etc) [1]. Some of the key advantages that NCS holds over traditional closed loop control system are reduced wiring (due to use of shared medium), increased agility of the system, low cost, modularity, flexibility regarding the data transmission architecture [2]. No wonder, NCS have found wide applicability in diverse fields like aircraft and space applications, factory automation, remote diagnostics etc. However, NCS is not without glitches. The introduction of the network within the loop, which brought in so many advantages, is as well, responsible for introducing some network related constraints which degrades the closed loop performance. Among these constraints, the first one is that of network induced delays i.e. the delay in transmission of data packets between different nodes of the control system. Secondly, there may be packet drop-outs (due to buffer overflows etc.) i.e. failure of the data packet to reach its intended recipient. Thirdly, the order of the packets may get changed during transmission and arrive at the recipient node in out-of-order sequence due to stochastic network induced delays with its upper magnitude greater than the sampling time of the system. These three constraints play pivotal roles in deciding the control loop performance over the network. Any one of the factors alone or more in unison can degrade the control performance of a well tuned loop. In worst case the system may become unstable altogether if these constraints are not taken into consideration during the controller design phase. Research groups from diverse backgrounds have worked to find ways to eliminate these factors all together or at least restrict their detrimental effect on the system performance. The current research activities in the field of NCS can be broadly classified into three domains. The first domain is the analysis and design of robust control systems that are capable of handling network induced irregularities [3]. Development of newer network transmission architecture, protocol and scheduling of events and revamping the existing ones in order to neutralize the ill effects of network induced delays and packet dropouts constitutes the second domain [4]. The third domain consists of optimal time domain tuning of PID controllers considering the random network induced delays and packet drop outs in NCS [5].

This multipronged research in the field of NCS, especially from the control system designing aspect, necessitates the requirement of a standard embeddable network test-bed to verify, in real-time, these proposed control algorithms [6]-[7]. Here, lies the motivation of the present work. This paper presents a detailed treatment on the design and implementation of an embeddable network test-bed which can be used as a platform to verify and validate real-time control algorithms in order to ensure optimal time domain performance in the presence of network induced delays and packet drop-outs even in the presence of out-of-order packets in the loop. These three features that essentially characterize the detrimental nature of a real-time transmission network, namely stochastic delay, data packet drop-out and arrival of packets out-of-order have been incorporated as a Stateflow based network test-bed. The settings for these parameters can be specified by the user. Hence, precise modeling of any given quality of network can be done with this test-bed to develop offline control algorithms that is capable of handling certain amount of stochastic delay, packet loss etc. Furthermore, a control algorithm consisting of a conventional PI controller, with optimal time-domain performance index based tuning via a stochastic algorithm i.e. Genetic Algorithm is tested out on the proposed network test-bed. The optimally tuned PI control algorithm has been considered as a test case and can also be easily extended to the other popular control methods to handle a specified level of network induced stochastic delays and drop-outs.

Manuscript received April 14, 2011. This work has been supported by the Board of Research in Nuclear Sciences (BRNS) of the Department of Atomic Energy (DAE), India, sanction no. 2009/36/62-BRNS, dated November 2009.

A. Mukherjee, S. Das, I. Pan and A. Gupta are with Department of Power Engineering, Jadavpur University, Salt Lake Campus, LB-8, Sector 3, Kolkata-700098, India (E-mail: ayanmukherjee.email@gmail.com).

A. Pakhira is with Department of Instrumentation & Electronics Engineering, Jadavpur University, Salt Lake Campus, LB-8, Sector 3, Kolkata-700098, India.

The closed loop time domain performances like a suitable error index and the control signal with the consideration of random network induced delays and packet drop-outs reflect stochastically varying non-smooth search space. Genetic algorithm has capability to carry out optimization more effectively in comparison with other gradient based methods in a stochastically varying search space. Thus in this paper, GA has been used as the stochastic evolutionary algorithm to tune the offline system in order to handle the unreliable network conditions effectively, as suggested by Pan *et al.* [5]. Integral of Time Multiplied Absolute Error (ITAE) has been taken as the time domain error index along with Integral of Squared Controller Output (ISCO) in order to keep the actuator size within limit and thereby preventing actuator saturation. In networked process control, the use of PI type controller is preferable since the derivative action of PID type controllers amplifies the randomness of the non-smooth stochastically varying objective function even upon optimization [5]. Thus derivative action in the controller structure is detrimental in NCS applications from actuator design point of view. The optimally tuned PI control algorithm is tested next over a real communication network to verify its suitability in real-time automation.

The rest of the paper is organized as follows. Section II describes the theoretical formulation which includes the basis for the process and tuning strategy selection etc. The design and modeling of the real-time network test-bed is detailed in Section III. Section IV elucidates the real-time implementation, results and discussions of the tuned control system performance over network. Section V contains the conclusions, followed by the references.

## II. Theoretical Formulation

### A. Design of PI Controller over NCS

A suitable choice of a test plant is important in order to test out the effectiveness of the optimal PI controller tuning strategy in the real-time network test-bed. The test plant considered for the present study is a FOPTD with balanced lag and delay type $(L = 1, T = 1.5, L \approx T)$ as has been studied Astrom & Hagglund [8] and is considered to be handled over communication network by a simple PI controller (2) having parallel structure.

$$P(s) = \frac{5}{1.5s+1}e^{-s} \tag{1}$$

$$C(s) = K_p + (K_i/s) \tag{2}$$

The rationale behind the selection of such a test plant is that such systems are hard to control. More so in the presence of the stochastic network induced delays and drop-outs, and out-of-order packets arrival. Hence the plant has to be well tuned using a carefully chosen optimization algorithm with proper consideration of the stochastic variation in network conditions to get a stable time response. Also the actual network conditions must be reproduced within the real-time network test-bed in order to test the control algorithms during the design phase.

### B. Time Domain Optimal Tuning of Networked Process Controllers via Stochastic Optimization

The optimal time domain tuning of the PI controllers (2) is done in order to stabilize the closed loop system in the presence of network induced irregularities relating to data packet transmission viz. random delay, packet drop and out of order packets. For optimal tuning of the PI controller a proper integral performance index needs to be chosen. Here, the control objective ($J$) consists of a weighted summation of ITAE and ISCO which is minimized with GA to get the optimal PI controller parameters. Here, the weights $w_1$ and $w_2$ balances the impact of error index and control effort and have been chosen to be same for simulation study.

$$J = \int_0^\infty \left[ w_1 \cdot t\left|e(t)\right| + w_2 \cdot u^2(t) \right] dt \tag{3}$$

The objective function becomes stochastic in nature in the presence of packet drop-outs and delays and can not be minimized by conventional gradient based optimization algorithms [5]. Hence GA is employed to find a near optimal solution within the specified search spaces of PI controller gains. Optimal controller design with stochastic optimization in NCS applications has been detailed in Pan *et al.* [5]. Whereas, the present paper focuses on the real-time network test-bed development to verify such control algorithms. The random delay, drop-out and out-of-order packets are implemented with Stateflow and its parameters can be pre-specified by the user. This is done to ensure the repeatability of the experiments and validate such control algorithms in a real-time controlled environment.

The PI parameters are tuned for a given probability of packet drop-out and stochastically varying delay with specified upper bound which is larger than the sampling time interval. Such a setting for stochastic delay propels the out-of-order packet arrival. In comparison to constant time delay, stochastic delays are much tougher to deal with [5] and can be efficiently stabilized by a non-gradient based optimization employed to minimize a stochastically varying fitness function (3).

## III. Modelling of the Real-Time NCS

The traditional PI controller is generally designed as a continuous time system. The introduction of communication network within the loop imparts a partial discrete characteristic to the system. Hence, NCS with continuous time plant and controller can be viewed as a hybrid system. Fig. 1 shows the schematic of the control loop closed over a real-time network. In the diagram, the controller-to-actuator delay is represented by $\tau^{CA}$ while $\tau^{SC}$ represents the sensor-to-controller delay. Both the delays are stochastically varying in nature. Additionally, there are dropped packets in the network due to buffer overflows, corrupt data etc.

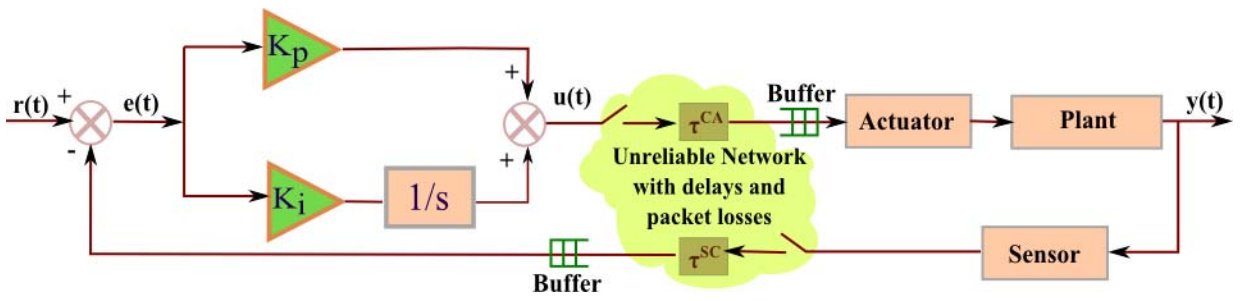

Fig. 1. NCS block diagram with PI Controller, discrete time network and continuous time plant.

While modelling the NCS, preservation of this hybrid trait is a must. In the present case, MATLAB/Simulink blocks have been used to model the continuous time plant and the PI type controller, which is widely used in process automation and controller design. The network has been simulated with the help of Stateflow charts, Simulink blocks and MATLAB functions that results in a flexible, modular utility. This design is capable of accurately mimicking the network delays (not limited to less than the sampling time like [9]) and packet drop-outs that essentially characterizes a real communication network.

Classical NCS literatures concerning analytical stability issues like Zhang *et al.* [9], Montestruque & Antsaklis [10] have considered network delays ( $\tau$ ) less than one sampling-time ( $T$ ). This is an over-simplified assumption and is often violated in practical NCS applications. A non-deterministic network condition often results in delays considerably larger than the sampling time of the NCS. Also, delay less than a sampling time rules out any possibility of out-of-order packets which is also not practical. In the present network modeling strategy, the occurrence of delays greater than one sampling time have been considered, along with packet-losses and out-of-order packet reception in order to enhance the controller design task in an optimization framework.

The network test-bed has been verified by sending continuous data packets and has been found to work reliably at the sampling rate used with no appreciable packet delays or drop-outs beyond specified limits. The network induced random delays are modeled, as mentioned above for both the forward loop (at the plant side) and the feedback loop (at the controller side). The presence of network induced stochastic delays has been restricted to the software-based simulation only to meet the control system designer's specification. This is to facilitate testing of a well behaved GA based near optimal control algorithm in a real-time NCS test-bed for reproducibility of the results.

### *A. Stateflow Based Real-Time Modeling of Random Delays, Packet Drops and Out-of-Order Packets in NCS*

Stateflow is an interactive, graphical design tool, based on finite-state machine theory that allows the user to develop and simulate event-driven systems [11]. It is tightly integrated with MATLAB and Simulink, providing the ability to include event-driven programming functionality within the Simulink models. A Stateflow chart can have input and output data (which serve as an interface to the Simulink environment), events for triggering the chart and actions and conditions which direct the flow of logic between the states. The Stateflow chart which is then available to the Simulink environment as another Simulink block can be connected to other Simulink blocks in a conventional way. Through these connections, the Stateflow and Simulink models share their data and respond to events.

Fig. 2 shows the overall network simulation model using Stateflow. The "Packet drop-out block" simulates the drop out of packets. The "Random delay" block is responsible for generating random stochastic delays in a specified probability distribution. The "Stateflow model for Out-of-order packet removal" block discards the out-of-order packets since they are highly detrimental from control system design point of view as in Zhao *et al.* [12]. Each Stateflow block except for "Stateflow model for out-of-order packet removal" block runs at a rate of 10 times higher than that of the packet generation rate in order to simulate delay with appreciable precision. "Stateflow model for out-of-order packet removal" block runs at the specified packet generation rate. New data is sampled in each of the Stateflow blocks at the end of 10 runs.

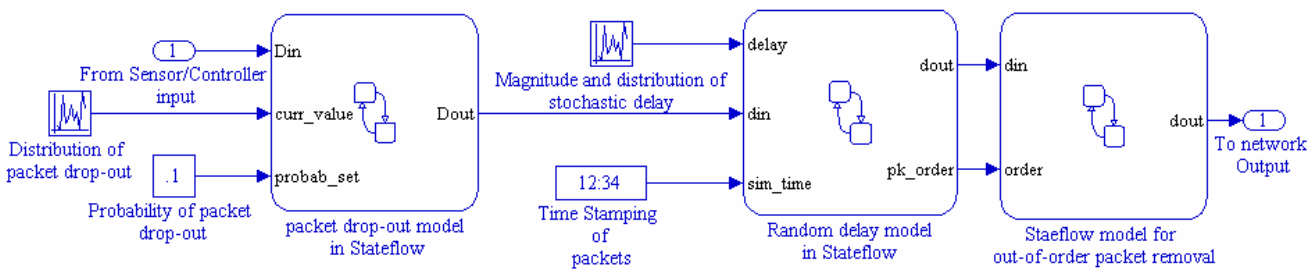

Fig. 2. Simulink & Stateflow based network test-bed.

As shown in Fig. 3 the "Packet drop-out model in Stateflow" block consists of a single state which calls an Embedded MATLAB function *drop()* after every 10 steps of the Stateflow chart. The function *drop()* determines whether an input will be accepted, based on the probability of packet drop which can be specified by the user.

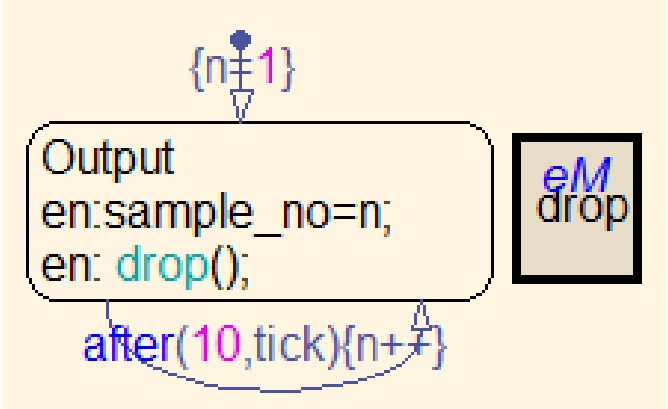

Fig.3. Stateflow model for simulating packet drop-out.

The "Random delay model in Stateflow" block, as shown in Fig. 4 consists of a single state and four Embedded MATLAB functions, viz. *init(), in_q(), sel_q() and out_q(n)*. Within this block, firstly the random delay is generated, the type and upper bound of which can be specified by the user. Then each incoming packet is held up by an amount of the time equal to its corresponding delay generated. The initialization of buffer and counter is implemented through *init()*. Function *in_q()* handles the data input and time stamping, *sel_q()* does the time-stamp comparison and *out_q()* outputs the data. The timing and order of execution of the Embedded MATLAB functions is controlled by the state Delay_and_pop.

The "Stateflow model for out-of-order packet removal" block, shown in Fig. 5, receives and stores (if needed) the time-stamped inputs (the last 1000 data packets in the present case). At every step, it outputs data corresponding to the highest value of time-stamping present in the buffer constrained to the condition that the present time stamp value should be greater than that of the last data passed through. Hence the older packets are sorted out and discarded.

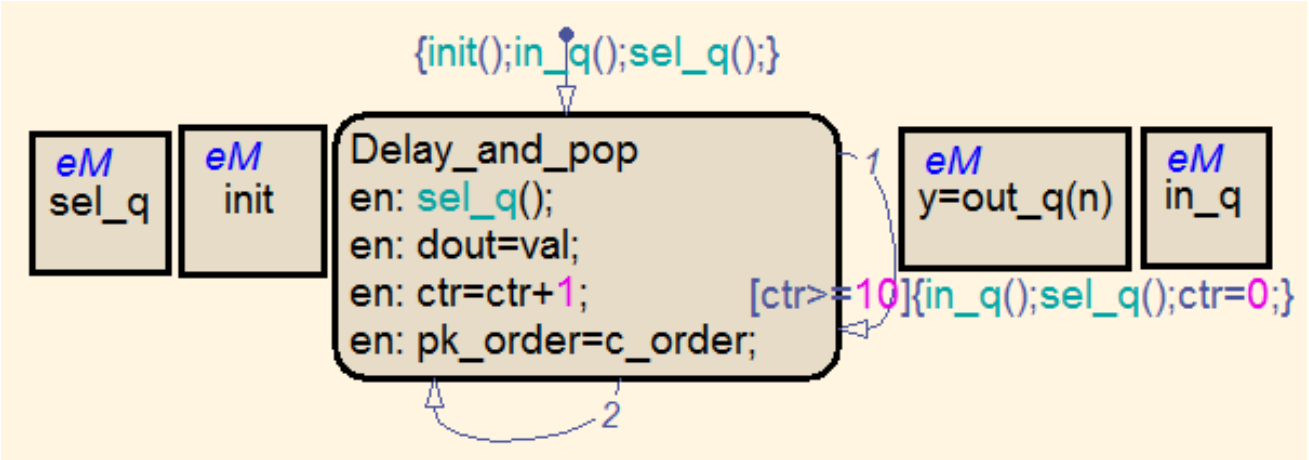

Fig. 4. Stateflow model for simulating random delay.

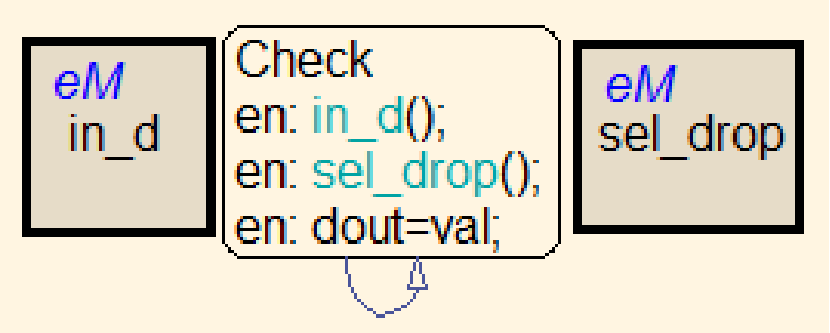

Fig. 5. Stateflow model for eliminating out-of-order packets.

*B. Hardware Setup, Synchronization between NCS Nodes and Its Necessity*

The hardware used to set up the embedded network test-bed consists of two Single Board Computers (SBCs) with embeddable kernels, a work-station and a 100 Mbps switched Ethernet. Switched Ethernet is a very effective communication medium for real-time system. Each SBC is an Advantech PCM-9572 board kernel based embedded system and acts as an xPC client. The work-station runs the MATLAB and Simulink based control algorithms and acts as the xPC host. The two SBCs running the xPC targets are used as the "Plant" and the "Controller" as shown in Fig. 6.

The transmission protocol chosen for real-time communication set-up is User Datagram Protocol/Internet Protocol (UDP/IP). Though otherwise perceived as an unreliable communication protocol, most of the features of UDP/IP like minimal error checking and no re-transmission ensures less network delay and data drop-out compared to other more secured transmission protocols and hence makes it very suitable for use in real-time-communication [13]. Hence with UDP/IP, data may be easily lost if the receiver fails to intercept it at the time of transmission. In view of the fact that both the SBCs are essentially sampled devices with such choice of transmission protocol, the sampling instances of both the plant-target and controller-target should be same. Failing this might lead to increased loss of data packets over and above that due to network conditions. Both the discrete-time sensor and the continuous-time plant and controller, receiving the sensor data through the network are time driven [14]. But being time driven with similar step size does not guarantee proper synchronization between the two nodes since the SBCs are physically separate devices and thus run on their own separate clocks. When a simulation is run in these devices on the xPC kernel, each simulation time step consist of a finite number of clock cycles. Since the two SBC clocks are not synchronized, the simulation time steps and consequently the time of packet transmission by the sender and polling for data by the receiver may not be at the exact same instance. Thus both the nodes need to be synchronized in order to preserve data integrity. This synchronism has been achieved by the use of interrupts and a separate connection between the plant and controller targets.

One of the xPC targets is considered to be the master device and the other as the slave. The data sending of the master xPC is allowed to run on its own clock-time and an interrupt signal is generated on its Parallel/Line Print Terminal (LPT) port at every time-step. The slave xPC receives those interrupts through its parallel port, then acknowledges the interrupt request and proceeds with its simulation. In the present case, the plant-node is made the master and the controller-node is made as the slave. Fig. 6 represents the schematic diagram about how the synchronization is made to work between the two xPC nodes.

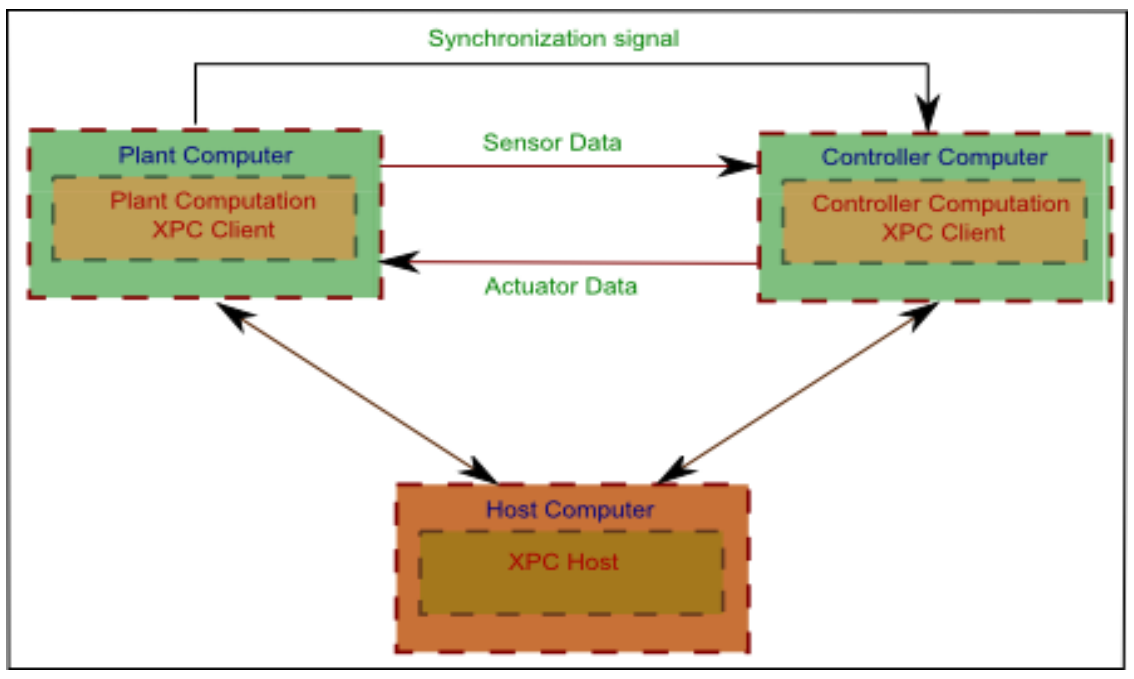

Fig. 6. Schematic diagram of synchronization between the two SBCs.

*C. Simulink Implementation of the NCS*

Stateflow models representing the network delays, packet drops etc. (in section IIIA) is used in the real-time NCS control algorithm test-bed along with the synchronization method as discussed in section IIIB, in the following manner:

*1) Simulation at Master (Plant) Side*

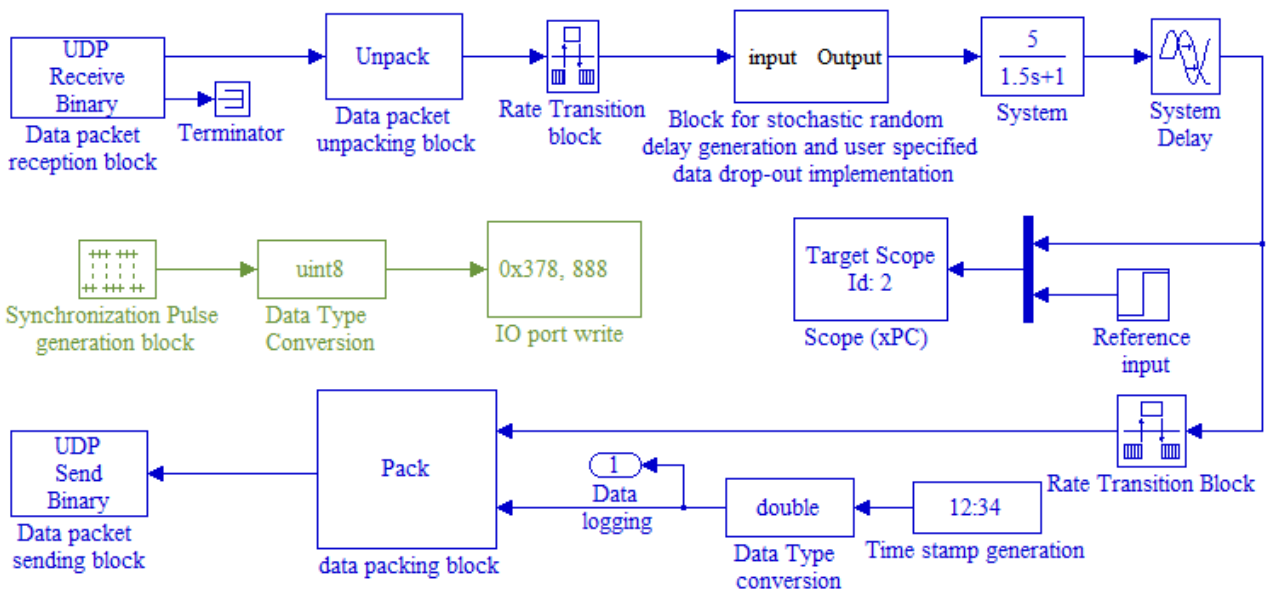

Fig. 7. Simulink model of Plant (Master) node.

Fig. 7 depicts the Simulink model developed for the master node. The "Synchronous pulse generation" block generates pulse signal at specified sampling instances which is then written to LPT port (Address = 0x378). The system block defines the plant model under consideration. The rest of the blocks are responsible for sampling plant output at specified time interval, packing-unpacking of out-going and in-coming data-streams respectively. Most importantly, the Stateflow based real-time network block mimics the specified network delay and drop-out characteristics.

### 2) Simulation at Slave (Controller) Side

The slave consists of an interrupt-triggered function call subsystem which handles the data transmission and reception. Whenever a rising edge is received in the LPT port data LSB, an interrupt is generated and is acknowledged by the "xPC Target IRQ Source" block which then triggers the function call subsystem. The PI controller is outside the subsystem. Data flow between the asynchronous (Synchronization pulse-triggered) and synchronous parts are handled by the "xPC Target Asynchcronous Rate Transition" or a combination of "Asynchronous Buffer Block Write" and "Asynchronous Buffer Block Read" blocks as shown in Fig. 8. This is to prevent data loss in the case when input or output to or from the triggered subsystem does not happen at the specific time-steps of the slave.

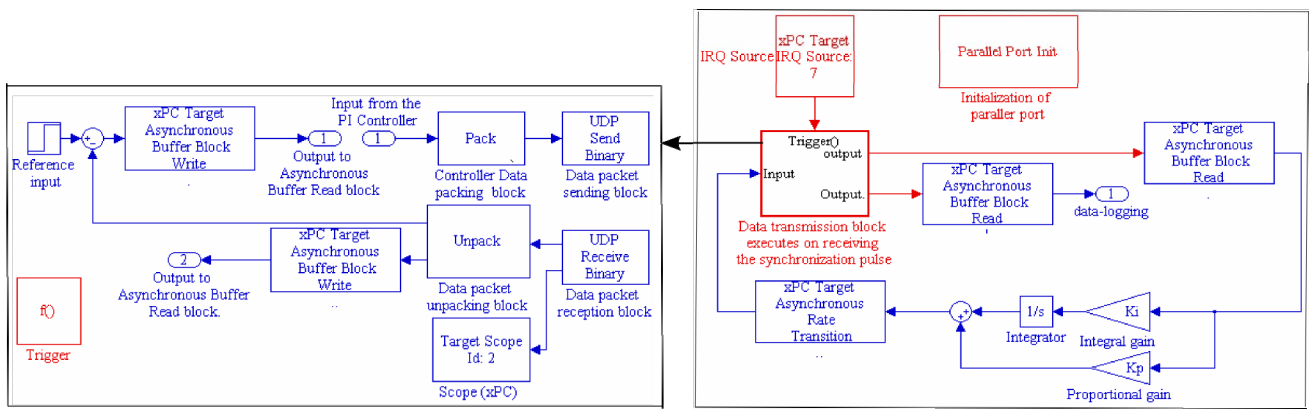

Fig. 8. Simulink model of Controller (Slave) node.

### D. Few Practical Implementation Issues

Previously, Simulink based delay simulation has been attempted using the "Variable Transport Delay" block in [5]. However this approach has resulted in all the packets getting shifted in time whenever a delay is present at an instant which does not fully mimic real network scenarios and does not allow packet level simulation. The use of the "Variable Time Delay" block results in non-deterministic data. Delayed data is not available at desired time instances as specified by the delay parameter.

The packet level network simulation (with random delays and drops) for controller design has been improved in [14] with the use of SimEvents blocks. SimEvents is a MATLAB package for Discrete Event Simulation (DES) that is extremely well-suited to simulate real networks. However, Simulink models containing SimEvents blocks are not embeddable in Real-time xPC targets. Hence this package was not usable for the present work. Stateflow overcomes all the difficulties faced using the afore-mentioned methods. It provides extremely flexible ways of manipulating data and program-flow and hence has successfully been used in the present work for simulating the delays with acceptable realism and accuracy. The delay generated is approximated to a value which is an integer multiple of the fixed step size. Any value of delay less than the fixed step size will fail to register itself as delay to the program. Thus, lower the fixed step size, finer will be the quantization level of the delay values. The fixed step size should not become comparable to the Task Execution Time (TET) which may lead to CPU overload. In the present case, a fine balance has been made that allows delay values to be generated with reasonable flexibility.

The present set-up does not include the scope to measure the extent of packet drop-out within the physical link itself. Thus before the actual implementation of the experimental set-up, the communication channel i.e. Ethernet link, in the present case is tested under various settings to find out the limits of different parameters and variables associated with the test-bed within which the Ethernet serves as a reliable communication channel.

### E. Data Logging and Its Advantage in Quantitative Performance Analysis in Real-Time NCS

Two monitors connected to the two SBCs shows the plant response and the corresponding controller output. But from these outputs only the qualitative nature of the responses can be judged and no further quantitative study can be made out of them. To overcome this limitation the data logging tool available with the xPC target set-up has been used. The data logging tool stores the specified signal values in actual runtime which can be retrieved later from the MATLAB Workspace for detailed analysis.

## IV. Simulations and Testing of the Real-Time PI Control Algorithm

For simulation study, the parameters adopted for optimal tuning (offline) of the PI parameters via GA are detailed next. The data packets are generated every 0.1 seconds. The probability of packet drop-out is set at 0.1 and the upper bound of network induced delay is taken to be 0.2 seconds. The control tuning algorithm had been run for a 30 seconds of time horizon. The optimally tuned PI controller parameters are $K_p = 0.1563$ and $K_i = 0.0939$ respectively with the corresponding expected minima of the stochastically varying objective function as $J_{\min} = 5.1713$.

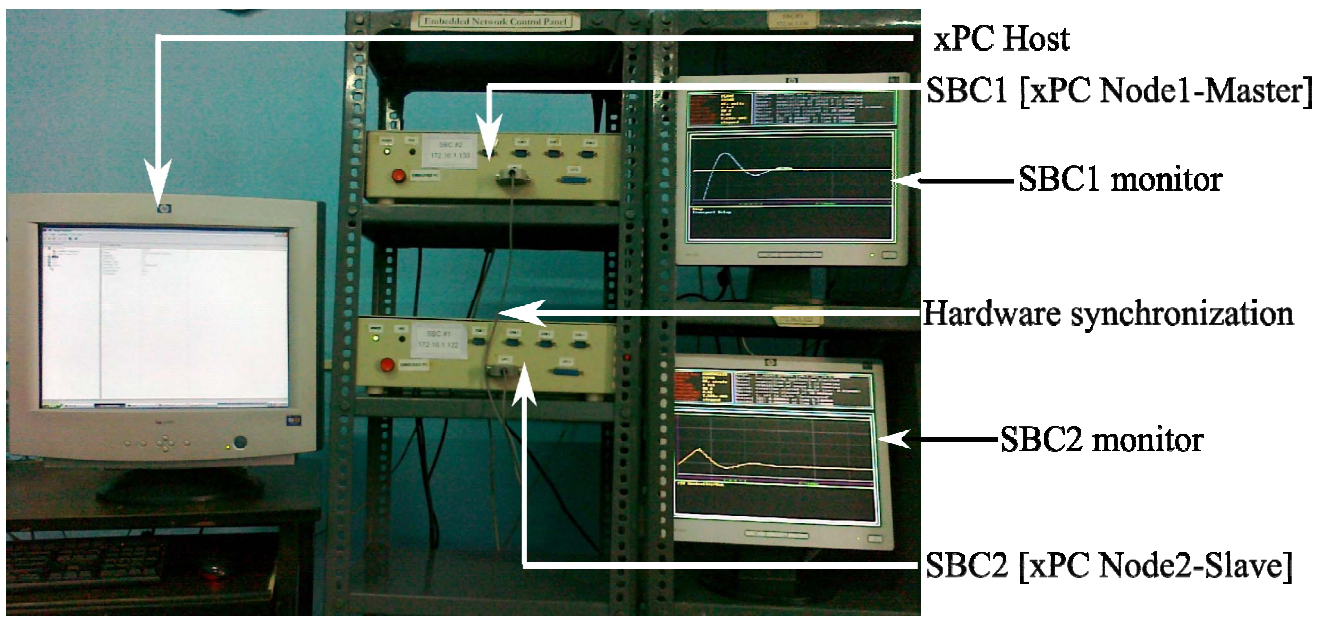

Fig. 9. Laboratory set-up for the real-time NCS test-bed.

Fig. 9 identifies the different hardware components of the NCS test-bed. The host xPC and the client xPCs (the two SBCs) are connected through 100 Mbps switched Ethernet. One SBC is assigned to act as the master node whereas the other one as the slave as also shown in the schematic Fig. 6. After establishing connection between the host and the client xPCs, the Simulink models of the two nodes are downloaded to their respective terminals. The two respective parallel ports are connected externally to ascertain proper synchronization between the nodes. The outputs of the plant and the controller are viewed on the respective monitors and logged in the computer memory. The plots are regenerated from the logged data as can be viewed in Fig. 10.

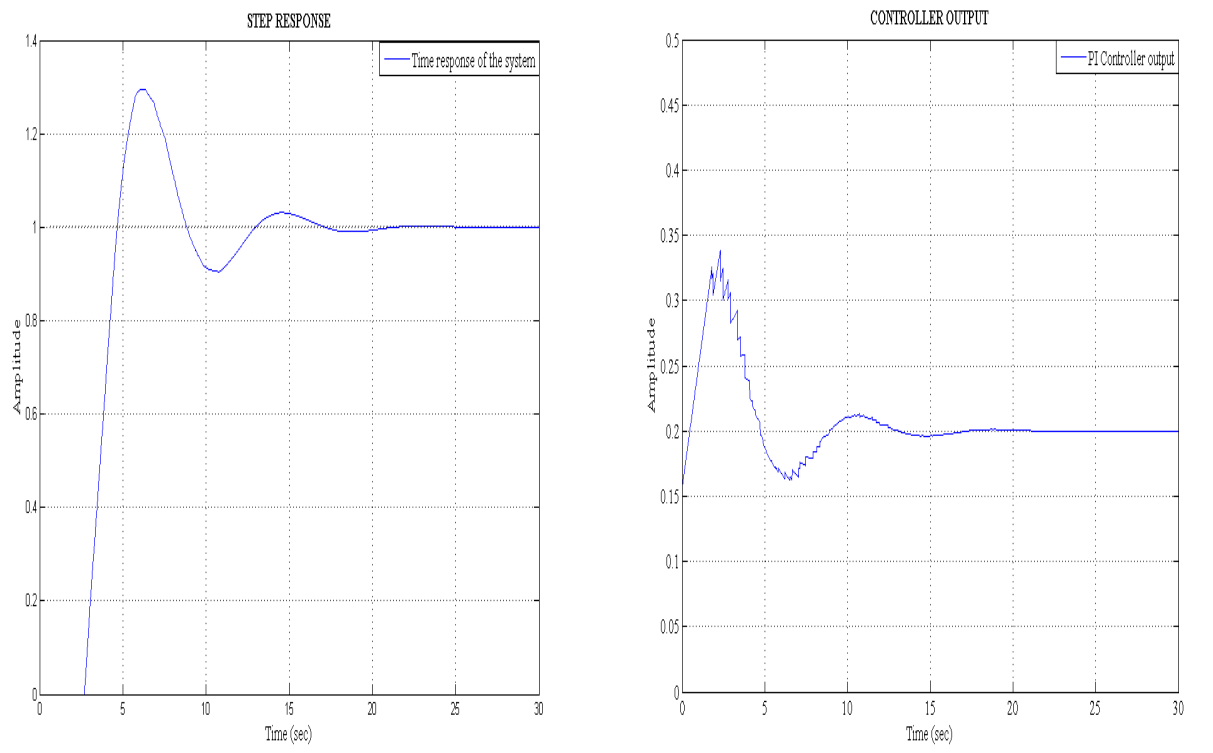

Fig. 10. Step response and control signal of the networked PI control system.

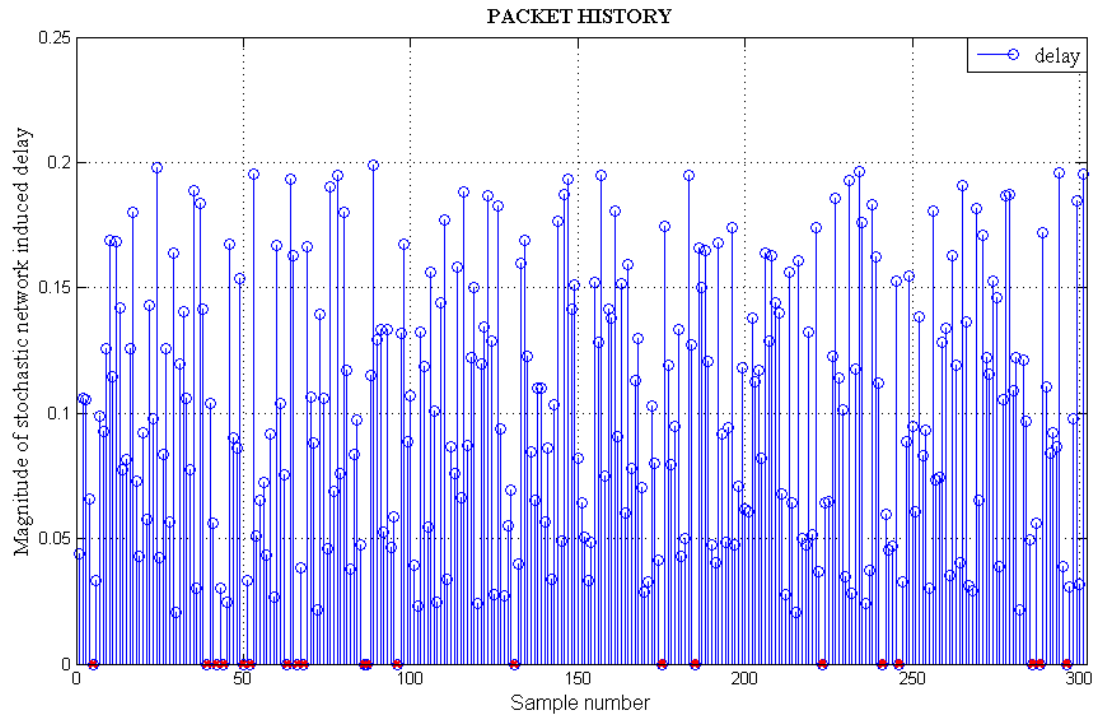

Fig. 12. Random delay of each packet and packet losses.

As shown in Pan *et al.* [5], an optimally tuned PID controller can be used in networked process control but suffers from large oscillation of control signal due to the randomness in the loop. The present methodology gives a smoother control action (Fig. 10) but the overshoot is slightly higher due to the absence of the derivative action as compared to [5]. Additionally, the magnitude of stochastic delay generated for each data packet as also the dropped packets are plotted in Fig.11. Here, the interspersed red circles represent the dropped packets and the blue stems correspond to the respective magnitude of random delays. From Fig. 10 it is evident that the plant, after being embedded in the proposed real-time network test-bed shows reasonably good closed loop time domain response using the optimally tuned PI controller parameters obtained from GA based tuning under network induced delays and drop-outs and filtering the order of data-packets.

## V. Conclusion

The new Stateflow based modeling of the network test-bed is validated with an optimal PI control system in the presence of random delays and drops. Simulation and experimental results on a real-time framework show the effectiveness of the proposed methodology. Future scope of research may be directed towards validating the network test-bed for different random delay distribution and more complicated frequency domain design of networked process controllers.